\documentclass[twocolumn,showpacs,preprintnumbers,amsmath,amssymb]{revtex4}

\usepackage{graphicx}
\usepackage{dcolumn}
\usepackage{bm}

\begin{document}
\title{
Evidence for Insulating Behavior in the Electric Conduction of 
(NH$_3$)K$_3$C$_{60}$ Systems 
}
\author{
H. Kitano$^1$, R. Matsuo$^1$, K. Miwa$^1$, A. Maeda$^1$, 
T. Takenobu$^{2,3}$, Y. Iwasa$^{2,3}$, T. Mitani$^2$
}
\address{
$^1$Department of Basic Science, The University of Tokyo, Meguro-ku, Tokyo 153-8902, Japan \\
$^2$Japan Advanced Institute of Science and Technology, Tatsunokuchi, Ishikawa 923-1292, Japan \\
$^3$Institute for Materials Research, Tohoku University, Aoba-ku, Sendai 980-8577, Japan
}
\date{}
\begin{abstract}
Microwave study using cavity perturbation technique revealed that 
the conductivity of antiferromagnet (NH$_3$)K$_{3-x}$Rb$_x$C$_{60}$ at 200K 
is already 3-4 orders of magnitude smaller than 
those of superconductors, K$_3$C$_{60}$ and (NH$_3$)$_x$NaRb$_2$C$_{60}$, 
and that the antiferromagnetic compounds are {\it insulators} below 250K 
without metal-insulator transitions.
The striking difference in the magnitude of the conductivity 
between these materials 
strongly suggests that 
the Mott-Hubbard transition in the ammoniated alkali fullerides 
is driven by a reduction of lattice symmetry 
from face-centered-cubic to face-centered-orthorhombic, 
rather than by the magnetic ordering. 
\end{abstract}

\maketitle

The electronic and magnetic properties in C$_{60}$ are 
the subjects of renewed interest\cite{Schon01,Makarova01}. 
In particular, the recent remarkable observations of 
high-temperature superconductivity (HTSC) in C$_{60}$ based field-effect 
transistors (C$_{60}$-FETs) \cite{Schon01} have attracted 
a lot of attention to non-cuprate HTSC. 
A critical temperature $T_c$ of 117~K for hole-doped C$_{60}$/CHBr$_3$ 
\cite{Schon01} seems to revive the possibility of exotic SC. 
On the other hand, remarkably, the BCS theory based on a band picture seems 
to work well even for the HTSC in C$_{60}$-FETs, since the simple relationship 
between $T_c$ and the effective volume $V$ per C$_{60}$, 
established in early studies on alkali($A$)-doped C$_{60}$ 
\cite{GunnarssonRMP97}, still holds for the HTSC in C$_{60}$-FETs. 
Thus, the understanding of the mechanism for SC in C$_{60}$ systems is 
the subject of urgent importance in the condensed matter physics. 

In close relation to this issue, recent studies on various alkali-doped 
fullerides strongly suggested the crucial roles of the electron-phonon (el-ph) 
and the electron-electron (el-el) interactions in C$_{60}$ systems. 
First, in contrast to band theory, 
$A_4$C$_{60}$ \cite{Murphy92} and Na$_2$C$_{60}$ \cite{Brouet01} 
have been reported to be nonmagnetic insulators, even though 
the latter material has a face-centered-cubic (fcc) structure 
that is isostructural to $A_3$C$_{60}$. 
Such a metal insulator transition (MIT) in $A_n$C$_{60}$ ($n$=2, 3, 4) 
may be understood in terms of a Jahn-Teller deformation of molecules 
\cite{Suzuki95,Han00}. 
The second example is seen in (NH$_3$)K$_3$C$_{60}$ 
which is isovalent to $A_3$C$_{60}$ but showed no SC \cite{Rosseinsky93}. 
Due to the insertion of a neutral ammonia molecule, 
the crystal structure is face-centered-orthorhombic (fco). 
Recent experiments on (NH$_3$)K$_3$C$_{60}$ demonstrated that 
the ground state is an antiferromagnetic (AF) insulator, 
suggesting an aspect of the Mott-Hubbard (MH) system 
\cite{Prassides99,Iwasa96,Simon00,Tou00}. 

The above two examples clearly indicate that the SC phase in C$_{60}$ systems 
competes with two kinds of insulating phases 
due to the strong el-ph and el-el interactions, 
suggesting the importance of both interactoins for the mechanism of HTSC. 
Thus, the full understanding of these instabilities in C$_{60}$ 
systems is crucially important. 

However, to establish the MH picture in (NH$_3$)K$_3$C$_{60}$ systems, 
a crucial question is whether the paramagnetic phase above 
the N\'{e}el temperature, $T_N$, is metallic or not. 
This issue is quite controversial. 
Some experiments concluded that the MIT occurred at $T_N$ ($\sim$40~K) 
\cite{Iwasa96,Simon00}, 
while others concluded that the high-temperature phase above $T_N$ 
was also insulating \cite{Tou00}. 
The main problem is that the previous experiments only investigated 
the magnetic properties of these materials 
\cite{Prassides99,Iwasa96,Simon00,Tou00,Takenobu00}. 
There has been no study on the electric conduction of them. 
This is mainly because these materials are obtained 
only in the powder form at present, 
and because the ammonia content is easily affected 
in the pelletizing processes. 
Because of these difficulties, 
the dc resistivity and the optical reflectivity have not yet been measured. 
A further complication seems to arise from 
the orientational order transition of the K-NH$_3$ pairs 
at $T_S$(=150~K) 
\cite{Ishii99}. 

In this paper, we report the first direct study of the electric conduction of 
(NH$_3$)K$_{3-x}$Rb$_x$C$_{60}$ ($x$=0, 2, 3) 
by using a microwave cavity perturbation technique. 
By comparing these results with our previous results from 
K$_3$C$_{60}$ and (NH$_3$)$_x$NaRb$_2$C$_{60}$ ($x$=0.8, 0.9) 
\cite{MaedaJJAP}, and also with the results from 
other reference powders such as Pb, V$_2$O$_3$, and C$_{60}$, 
we concluded that the electric conduction of (NH$_3$)K$_{3-x}$Rb$_x$C$_{60}$ 
is {\it insulating} between 4.5~K and 250~K, without any MIT below 250~K. 
These results are in sharp contrast to the metallic nature 
of superconducting (NH$_3$)$_x$NaRb$_2$C$_{60}$ \cite{MaedaJJAP}. 
The striking difference between both materials strongly indicates 
that the Mott-Hubbard transition (MHT) in the ammoniated alkali fullerides is 
driven by a reduction of lattice symmetry from fcc to fco, 
rather than by the magnetic ordering. 

The preparation of fco (NH$_3$)K$_{3-x}$Rb$_x$C$_{60}$ and 
fcc (NH$_3$)$_x$NaRb$_2$C$_{60}$ compounds 
has already been reported elsewhere \cite{Takenobu00,Shimoda96}. 
All sample powders were sealed in glass tubes under He. 
The average diameter of sample powders were estimated as 10$\pm$7~$\mu$m 
by the observation of the microscope image \cite{MaedaJJAP}. 
To measure the response for microwave 
magnetic ($H_\omega$) and electric ($E_\omega$) fields, 
we prepared a copper cylindrical cavity resonator operating at 10.7~GHz in 
the TE$_{011}$ mode, so one can choose the sample position 
between the antinode of the $H_\omega$-field and that of the $E_\omega$-field. 
The microwave loss $\Delta(1/2Q)$ and the frequency shift $\Delta f/f$ were 
measured between 4.5~K and 250~K. 
The response of the sample was obtained by subtracting the contribution 
of an empty tube with almost the same size as the sample tube. 
A set of careful measurements 
confirmed that 
the contribution of the glass tube to $\Delta (1/2Q)$ was very small, 
although the magnitude of $\Delta f/f$ 
was difficult to determine precisely. 
More details were described elsewhere \cite{MaedaJJAP}. 
\begin{figure}[b]
  \begin{center}
    \includegraphics[width=8.0cm]{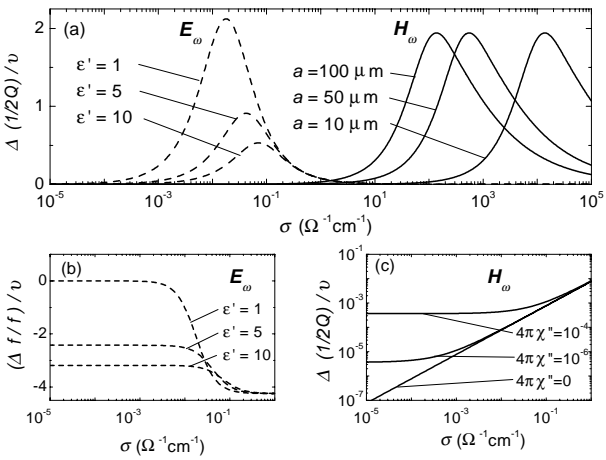}
     \caption{(a) Calculated $\Delta (1/2Q)$ at $E_\omega$ and $H_\omega$ 
     as functions of $\sigma$, for some values of $\epsilon'$ and $a$. 
     Here, $\omega$ is 2$\pi\times$10.7~GHz. 
     (b) Calculated $(\Delta f/f)_E$ for some values of $\epsilon'$.
     (c) $\Delta (1/2Q)_H$ for some values of $\chi"$. 
     }
    \label{fig1}
  \end{center}
\end{figure}

Our idea to study the electric conduction of unknown powders is quite simple. 
$\Delta (1/2Q)$ and $\Delta f/f$ at $H_\omega$ (or $E_\omega)$ are 
usually given as functions of the complex dielecric constant, 
$\epsilon$(=$\epsilon'+i\epsilon''$), 
the complex magnetic susceptibility, $\chi$(=$\chi'+i\chi''$), 
and the sample size $a$ \cite{Klein93}. 
In Fig.~1(a), we plot the calculated $\Delta (1/2Q)$ at $H_\omega$ 
and at $E_\omega$ [$\Delta (1/2Q)_H$ and $\Delta (1/2Q)_E$, respectively] 
as a function of the conductivity $\sigma$(=$\omega\epsilon''/4\pi$) 
for several values of $a$ and $\epsilon'$. 
All data of $\Delta (1/2Q)$ are normalized by $\upsilon$, 
which is the volume ratio of the sample to the cavity. 
As shown in Fig.~1(a), we found that $\Delta (1/2Q)_H\gg\Delta (1/2Q)_E$ 
in the high conductive region ($\sigma$$\gtrsim$~100~$\Omega^{-1}$cm$^{-1}$), 
independent of $a$ and $\epsilon'$, while $\Delta (1/2Q)_H\ll\Delta (1/2Q)_E$ 
in the low conductive region ($\sigma$$\lesssim$~0.1~$\Omega^{-1}$cm$^{-1}$), 
for a nonmagnetic material ($\chi''$$\sim$0). 
Thus, we can determine whether an unknown sample is conductive or not, 
only by comparing $\Delta (1/2Q)_H$ with $\Delta (1/2Q)_E$. 
Furthermore, as shown in Fig.~1(b), 
since $(\Delta f/f)_E$ increases with decreasing $\sigma$ 
we can determine the sign of $d\sigma/dT$ from 
the $T$ dependence of $(\Delta f/f)_E$, as discussed below. 

For a magnetic insulator with a large $\chi''$, 
the situation is not so simple. 
As shown in Fig.~1(c), 
$\Delta (1/2Q)_H$ saturates with decreasing 
$\sigma$ when $\chi''$ is finite. 
In such a case, $\Delta (1/2Q)_H$ is dominated by $\chi''$, 
while $\Delta (1/2Q)_E$ depends on $\sigma$ and $\epsilon'$. 
Thus, $\Delta (1/2Q)_H$ and $\Delta (1/2Q)_E$ are expected to behave 
quite differently from each other. 
\begin{figure}[b]
  \begin{center}
    \includegraphics[width=8.0cm]{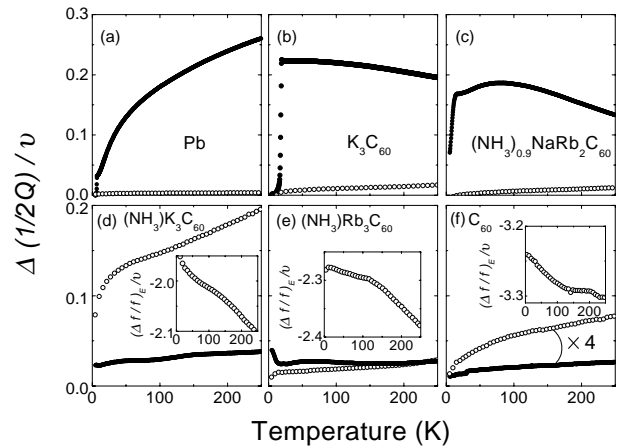}
     \caption{The measured data of $\Delta (1/2Q)$ for 
     (a)~Pb, (b)~K$_3$C$_{60}$, (c)~(NH$_3$)$_{0.9}$NaRb$_2$C$_{60}$, 
     (d)~(NH$_3$)K$_3$C$_{60}$, (e)~(NH$_3$)Rb$_3$C$_{60}$, and (f)~C$_{60}$, 
     respectively. Solid symbols are $\Delta (1/2Q)$ at $H_\omega$, and 
     open symbols are $\Delta (1/2Q)$ at $E_\omega$. 
     Insets: $(\Delta f/f)$ at $E_\omega$ for each material.}
    \label{fig2}
  \end{center}
\end{figure}

Figure~2 shows $\Delta (1/2Q)_H$ and $\Delta (1/2Q)_E$ as functions of $T$, 
for several materials including (NH$_3$)$A_3$C$_{60}$. 
We also plot $(\Delta f/f)_E$ in the insets of the lower panels. 
First, we confirmed that $\Delta (1/2Q)_H\gg\Delta (1/2Q)_E$ for Pb, 
as a reference of metals, and that 
$\Delta (1/2Q)_H\ll\Delta (1/2Q)_E$ for C$_{60}$, 
as a reference of insulators, 
as shown in Figs.~2(a) and 2(f), respectively. 
We found that $\Delta (1/2Q)_H\gg\Delta (1/2Q)_E$ 
for K$_3$C$_{60}$ and (NH$_3$)$_{0.9}$NaRb$_2$C$_{60}$ from $T_c$ to 250~K. 
These results provide direct evidence for the metallic nature in 
$A_3$C$_{60}$ showing SC,
even if $T_c$ is strongly reduced by the intercalation of NH$_3$. 

We found that $\Delta (1/2Q)_H$ for Pb decreased with decreasing $T$, 
while that for K$_3$C$_{60}$ and (NH$_3$)$_x$NaRb$_2$C$_{60}$ increased 
with decreasing $T$, as shown in Figs.~2(a) to 2(c). 
However, such a behavior depends on $\sigma$ and $a$. 
Typical value of $\sigma$ is 
$\sim$10$^3$~$\Omega^{-1}$cm$^{-1}$ for K$_3$C$_{60}$ ($T$=$T_c$) 
\cite{GunnarssonRMP97}, 
and is $\sim$10$^5$~$\Omega^{-1}$cm$^{-1}$ for Pb ($T$=77~K) 
\cite{AshcroftMermin}. 
On the other hand, the diameter of K$_3$C$_{60}$ powders was 
$\sim$10~$\mu$m \cite{MaedaJJAP}, 
while that of Pb powders was 90$\sim$125~$\mu$m. 
Thus, $\Delta (1/2Q)_H$ for Pb (or K$_3$C$_{60}$) decreases (or increases) 
with increasing $\sigma$, 
as predicted in Fig.~1(a). 

A quite different result was obtained for (NH$_3$)K$_3$C$_{60}$. 
As shown in Fig.~2(d) and its inset, 
the results of (NH$_3$)K$_3$C$_{60}$ indicated 
that $\Delta (1/2Q)_E\gg\Delta (1/2Q)_H$ between 4.5~K and 250~K, 
and that $(\Delta f/f)_E$ increased with decreasing $T$, 
which were quite similar to the case of C$_{60}$. 
These results strongly suggest that the electric conduction of 
(NH$_3$)K$_3$C$_{60}$ was {\it insulating} 
over the entire temperature range measured. 
\begin{figure}[h]
  \begin{center}
    \includegraphics[width=8.0cm]{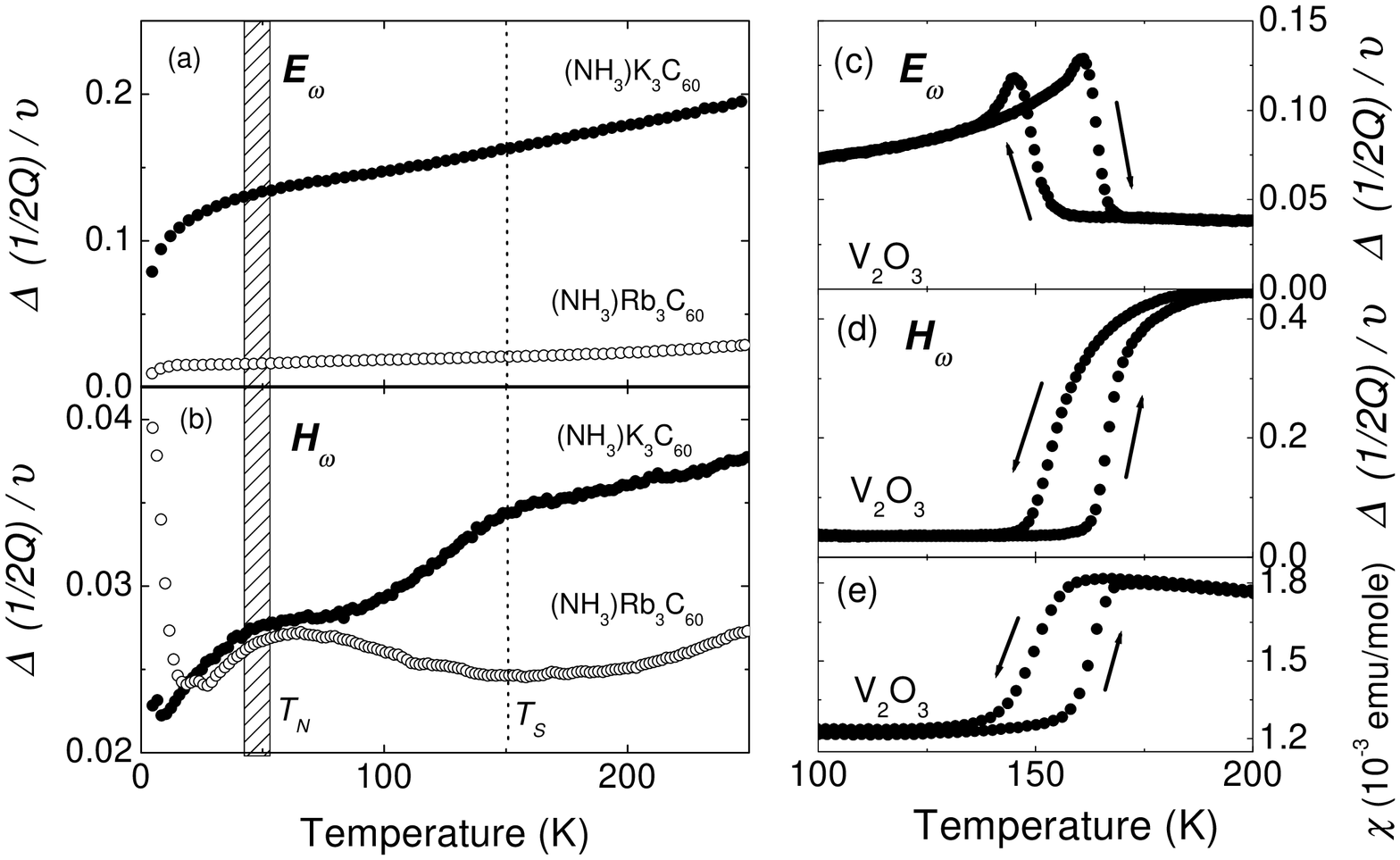}
     \caption{(a) $\Delta (1/2Q)_E$ for 
     (NH$_3$)K$_3$C$_{60}$ and (NH$_3$)Rb$_3$C$_{60}$. 
     (b) $\Delta (1/2Q)_H$ for (NH$_3$)K$_3$C$_{60}$ and (NH$_3$)Rb$_3$C$_{60}$. 
     (c) $\Delta (1/2Q)_E$ for V$_2$O$_3$.
     (d) $\Delta (1/2Q)_H$ for V$_2$O$_3$.
     (e) the dc magnetic susceptibility $\chi(T)$ of V$_2$O$_3$ 
     The contribution of impurities was subtracted.}
    \label{fig3}
  \end{center}
\end{figure}

The result of (NH$_3$)Rb$_3$C$_{60}$ depicted in Fig.~2(e) 
was slightly different from that of (NH$_3$)K$_3$C$_{60}$. 
That is, $\Delta (1/2Q)_E$$\approx$$\Delta (1/2Q)_H$ 
between 4.5~K and 250~K, which is different from 
those of Pb and C$_{60}$. 
However, this does not mean that $\sigma$ of (NH$_3$)Rb$_3$C$_{60}$ 
is larger than that of (NH$_3$)K$_3$C$_{60}$. 
Since the behavior of $(\Delta f/f)_E$ 
of (NH$_3$)Rb$_3$C$_{60}$ was similar to 
that of (NH$_3$)K$_3$C$_{60}$ and of C$_{60}$, 
it is suggested that $\sigma$ of (NH$_3$)Rb$_3$C$_{60}$ 
also decreased with decreasing $T$, 
and that it was located in a lower conductive region than 
the peak of $\Delta(1/2Q)_E$ in Fig.~1(a). 
Indeed, $\Delta (1/2Q)_E$ of (NH$_3$)Rb$_3$C$_{60}$ was found to be 
nearly an order of magnitude smaller than that of (NH$_3$)K$_3$C$_{60}$, 
while $\Delta (1/2Q)_H$ for both compounds were almost the same. 
We also found that the behavior of $\Delta (1/2Q)_E$ and 
$\Delta (1/2Q)_H$ were quite different from each other, 
as shown in Figs.~3(a) and 3(b). 
These features strongly suggest that $\Delta (1/2Q)_H$ of 
(NH$_3$)K$_3$C$_{60}$ and (NH$_3$)Rb$_3$C$_{60}$ 
were governed by $\chi''$ 
, as was already discussed. 
Thus, we conclude that $\sigma$ of (NH$_3$)Rb$_3$C$_{60}$ was 
also {\it insulating}, being smaller than 
that of (NH$_3$)K$_3$C$_{60}$. 

Next, we discuss the controversial issue whether or not 
the MIT occurs in (NH$_3$)K$_3$C$_{60}$ systems with varing temperature. 
For this purpose, we also measured V$_2$O$_3$ powders. 
V$_2$O$_3$ is a typical Mott insulator which exhibits a MIT at 
$T_{\rm MI}$=150$\sim$160~K \cite{McWhan69}. 
In Fig.~3, we compare the results of (NH$_3$)K$_{3-x}$Rb$_x$C$_{60}$ 
with those of V$_2$O$_3$. 
As shown in Figs.~3(c) and 3(d), we found that 
both $\Delta (1/2Q)_E$ and $\Delta (1/2Q)_H$ 
changed significantly at $T_{\rm MI}$. 
Here, $T_{\rm MI}$ was determined by measurement of 
the magnetic susceptibility using a dc SQUID magnetometer, independently, 
as shown in Fig.~3(e). 
Furthermore, we observed that 
$\Delta (1/2Q)_H\gg\Delta (1/2Q)_E$ above $T_{\rm MI}$ 
while $\Delta (1/2Q)_H\ll\Delta (1/2Q)_E$ below $T_{\rm MI}$, 
as was predicted in Fig.~1. 
These results indicate that both $\Delta (1/2Q)_E$ and $\Delta (1/2Q)_H$ 
are sensitive probes of the MIT. 
However, Figs.~3(a) and 3(b) show that $\Delta (1/2Q)_E$ 
varied very smoothly from 4.5~K to 250~K for both 
(NH$_3$)K$_3$C$_{60}$ and (NH$_3$)Rb$_3$C$_{60}$. 
Although $\Delta (1/2Q)_H$ of both compounds appeared to change slightly 
near $T_N$ or $T_S$, no anomaly was observed in $\Delta (1/2Q)_E$ 
at $T_N$ and $T_S$. 
Thus, we concluded that the MIT did not occur in (NH$_3$)K$_3$C$_{60}$ systems 
over the whole temperature range measured. 

The above discussion based on $\Delta (1/2Q)$ and $\Delta f/f$ is 
straightforward but somewhat qualitative, 
and should be confirmed in terms of $\sigma$. 
We tried to estimate $\sigma$ as follows. 
First, the complex dielectric constant in the powder form, $\epsilon_p$, 
was obtained from $\Delta (1/2Q)$ and $\Delta f/f$ at $E_\omega$, 
by using the formula in the so-called ``depolarization regime ", 
where the $E_\omega$-field almost uniformly penetrates into the sample 
\cite{Klein93}. 
\begin{equation}
\label{eq1}
\left(\frac{\Delta f}{f}\right)_E
-i\Delta\left(\displaystyle\frac{1}{2Q}\right)_E=
-\frac{\gamma}{n}\frac{\epsilon_p-1}{\epsilon_p-1+\displaystyle\frac{1}{n}}, 
\end{equation}
\noindent
where $\gamma$ and $n$ is the geometrical factor ($\propto\upsilon$) 
and the depolarization factor (typically, $n$$\sim$0.4), respectively. 
Next, the effect of the powder form (porosity) was corrected by using 
the so-called B\"{o}ttcher formula \cite{Bottcher}, 
\begin{equation}
\label{eq2}
\frac{\epsilon-1}{\epsilon+2\epsilon_p}=
\frac{1}{\delta}\frac{\epsilon_p-1}{3\epsilon_p},
\end{equation}
\noindent
where $\delta$, $\epsilon$ are the packing fraction of the sample powder, 
and the complex dielectric constant of the bulk sample, respectively. 
We estimated $\delta$ as 0.2$\sim$0.25 for almost all samples, 
by comparing the apparent volume packed in the glass tube 
with the true volume estimated from the specific gravity \cite{Kitano}. 
Finally, $\sigma$ was obtained from $\epsilon''$(=$4\pi\sigma/\omega$). 
In practice, the ambiguity in $\Delta f/f$ made 
the precise estimate of $\epsilon$ quite difficult. 
To avoid this difficulty, we utilized the fact that $\epsilon_p'$ depended 
on $\epsilon'$ only very weakly when $\delta$ was small. 
We found that $\epsilon_p'$ calculated by Eq.~(\ref{eq2}) was only varied 
from 1 to 2 for $\epsilon'$ ranged from 1 to 10. 
Thus, we assumed that 
$\epsilon'$ of (NH$_3$)K$_{3-x}$Rb$_x$C$_{60}$ was roughly 4$\sim$10 
($\epsilon'$$\sim$4 for C$_{60}$ \cite{Alers92}), 
and added a constant to $\Delta f/f$ so that $\epsilon_p'$=1$\sim$2. 

Figure~4(a) shows $\sigma$ estimated at 200~K in this way for 
(NH$_3$)K$_{3-x}$Rb$_x$C$_{60}$ ($x$=0,2,3). 
In the same figure, we also plot our previous results for 
K$_3$C$_{60}$ and (NH$_3$)$_x$NaRb$_2$C$_{60}$ ($x$=0.8,0.9) 
\cite{MaedaJJAP}. 
We estimated the error bars for $\sigma$, considering the ambiguity of 
$\epsilon'$(=4$\sim$10) and $\delta$(=0.2$\sim$0.25) for 
(NH$_3$)K$_{3-x}$Rb$_x$C$_{60}$, and that of $a$(=1$\sim$50~$\mu$m) for 
K$_3$C$_{60}$ and (NH$_3$)$_x$NaRb$_2$C$_{60}$. 
In spite of fairy large error bars, Fig.~4(a) clearly shows 
that $\sigma$ of (NH$_3$)K$_{3-x}$Rb$_x$C$_{60}$ 
is already at least 4 orders of magnitude smaller than 
the Mott limit for C$_{60}$ systems 
($\sigma_{\rm Mott}$=500$\sim$700~$\Omega^{-1}$cm$^{-1}$)\cite{Palstra94}. 
Thus, the results of Fig.~4(a) give quantitative support to our conclusion 
that (NH$_3$)K$_{3-x}$Rb$_x$C$_{60}$ is a genuine insulator. 
\begin{figure}[t]
  \begin{center}
    \includegraphics[width=8.0cm]{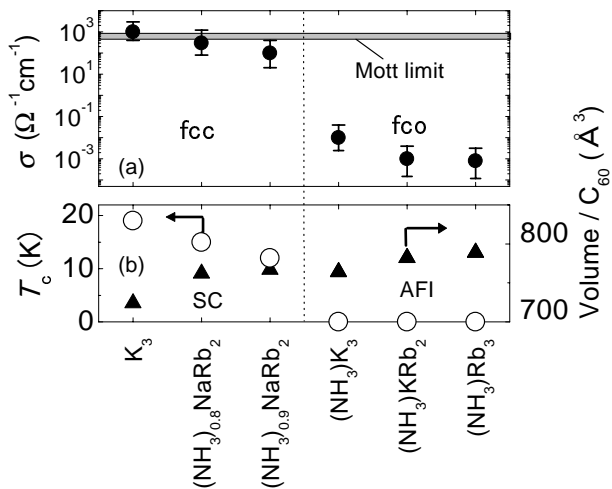}
     \caption{(a) The estimated values of $\sigma$ at 200~K for 
     K$_3$C$_{60}$, 
     (NH$_3$)$_{0.8}$NaRb$_2$C$_{60}$, 
     (NH$_3$)$_{0.9}$NaRb$_2$C$_{60}$, 
     (NH$_3$)K$_3$C$_{60}$, 
     (NH$_3$)KRb$_2$C$_{60}$, 
     (NH$_3$)Rb$_3$C$_{60}$, respectively. 
     (b) The plots of $T_c$ (open circle), and the volume $V$ per C$_{60}$ 
     (solid triangle), respectively.}
    \label{fig4}
  \end{center}
\end{figure}

When we compare these results with the previous results, 
our conclusion agrees with 
the recent NMR result by Tou {\it et al.} \cite{Tou00}, 
but differs from the others \cite{Iwasa96,Simon00}. 
We found that a sample tube of (NH$_3$)K$_{3-x}$Rb$_x$C$_{60}$ 
including a small amount of the residual SC phase 
showed an apparent metallic behavior similar to 
(NH$_3$)$_x$NaRb$_2$C$_{60}$. 
One possible reason for the controversy may be such a non negligible 
contribution of the residual metallicity. 


In Fig.~4(b), we plot $T_c$ and $V$, respectively. 
When we compare the results of Fig.~4(a) with 4(b), 
it is strongly suggested that the disappearance of SC 
is closely related to the drastic change in $\sigma$, 
implying the occurrence of MIT. 
This is most likely understood in terms of MHT, 
since the insulating phase was an antiferromagnet 
\cite{Iwasa96,Prassides99,Simon00,Tou00,Takenobu00}. 

What is the origin of this MHT ? 
We must note that MHT does not occur with varing temperature, 
in contrast to the previous conclusion \cite{Iwasa96,Simon00}. 
As shown in Fig.~4(b), it is also unrelated to the change of $V$. 
We consider that the most probable candidate is 
a reduction of lattice symmetry from fcc to fco. 
A recent theoretical study \cite{Han00} has suggested 
that the fcc structure of $A_3$C$_{60}$ favors 
the larger critical value $(U/W)_c$ for which the MHT occurs. 
Another possibility is that the less-symmetric configuration of C$_{60}$ 
removes the degeneracy of the $t_{1u}$ band, 
leading to a decrease of $(U/W)_c$ \cite{GunnarssonRMP97}. 
Such a MHT due to the lattice-symmetry reduction strongly suggests 
the significance of the highly symmetric configuration of C$_{60}$ 
for the bulk SC in $A_3$C$_{60}$. 

Interestingly, in the cases of C$_{60}$-FETs \cite{Schon01}, 
the lattice-symmetry reduction seems to be unimportant. 
Although the reason is unknown at present, 
such insensitivity of SC to the lattice symmetry 
may be related to the SC occuring only in a single layer of C$_{60}$ crystal. 
This deserves further investigation with prior importance. 

In conclusion, we studied the electric conduction of 
(NH$_3$)K$_{3-x}$Rb$_x$C$_{60}$, by using the cavity perturbation technique. 
We confirmed that (NH$_3)$K$_{3-x}$Rb$_x$C$_{60}$ was 
insulating between 4.5~K and 250~K, without any MIT at $T_N$ and at $T_S$. 
We also found that $\sigma$ of (NH$_3$)K$_{3-x}$Rb$_x$C$_{60}$ at 200~K 
was already 3-4 orders of magnitude smaller than 
those of K$_3$C$_{60}$ and (NH$_3$)$_x$NaRb$_2$C$_{60}$. 
From the striking difference in $\sigma$ between these materials, 
we conjecture that the MHT in $A_3$C$_{60}$ systems is driven by 
the reduction of lattice symmetry from fcc to fco. 

We thank T. Umetsu, R. Inoue, and H. Shimoda for experimental help, 
and D. G. Steel for a critical reading of the manuscript. 
This work was partly supported by the Grant-in-Aid for Scientific Research 
(10184101, 12046222, 13440110 and 13750005) 
from the Ministry of Education, Science, Sports and Culture of Japan. 


%
%
%
%

\end{document}